# Measurement of the orbital and superhump periods of the eclipsing cataclysmic variable SDSS J170213.26+322954.1

David Boyd, Arto Oksanen & Arne Henden

The orbital period of the eclipsing CV SDSS J170213.26+322954.1 has been measured as 0.10008215±0.00000001d using observations of eclipses during the first recorded superoutburst in 2005 October, together with eclipses observed in quiescence in 2003 July. This period puts the system in the centre of the period gap. Observation of superhumps during the 2005 October outburst with a period of 0.10496±0.00015d confirms this to be a UGSU-type system with a period excess of 4.9%.

## Discovery

This cataclysmic variable star (CV) was first identified spectroscopically in the Sloan Digital Sky Survey database.[1] Follow-up photometric observations of the system (hereafter referred to as SDSSJ1702) in quiescence were obtained by Henden at the US Naval Observatory's Flagstaff Station (NOFS) on 2003 July 3 using a 1.0m reflector. These showed it to be an eclipsing system with a period of approximately 2.5 hrs. The lightcurve is shown in Figure 1.

**Table 1.** Observations of SDSSJ1702 during 2005 October

| Date (2005) | Range of JD 2453600+ | Mean mag (excl.eclipses) | No. of eclipses | Filter | Observer |
|---|---|---|---|---|---|
| Oct 04 | 48.304–48.454 | 14.1 | 2 | C | DB |
| 05 | 49.237–49.385 | 14.6 | 2 | C | AO |
| 06 | 50.212–50.384 | 14.5 | 2 | C | AO |
| 09 | 53.238–53.358 | 14.9 | 2 | C | AO |
| 09 | 53.289–53.438 | 14.8 | 1 | C | DB |
| 10 | 54.243–54.380 | 15.1 | 1 | C | AO |
| 10 | 54.301–54.408 | 15.0 | 1 | C | DB |
| 11 | 55.227–55.373 | 15.4 | 2 | C, R | AO |

UCAC-2 gives the position of the variable as RA 17h 02m 13.26±0.13s, Dec +32d 29m 54.08±0.08s.

## First detected outburst

SDSSJ1702 was detected in outburst for the first time at magnitude 13.7 by Patrick Schmeer on 2005 September 30 and reported on October 3 after confirmation by Tonny Vanmunster.[2] Time series unfiltered photometry was carried out by David Boyd (DB) on 3 nights starting on October 4 using a 0.35m SCT and by Arto Oksanen (AO) on 5 nights starting on October 5 using a 0.4m SCT. Comparison star magnitudes were provided by Henden.[3] During the outburst 13 eclipses were recorded. All data were obtained unfiltered (C) except for October 11 when the second eclipse was observed through an R filter. By October 13 the magnitude had dropped to 16.5 and the outburst was effectively over. Details of the observations are given in Table 1. Examples of lightcurves are shown in Figures 2 and 3.

The field of SDSSJ1702 is shown in Figure 4. Astrometry using *Astrometrica*[4] and

## Orbital period

All times were heliocentrically corrected and times of minimum were measured for each eclipse using two methods, a quadratic fit developed by Boyd and the 5-term Fourier fit in Nelson's *Minima* software package.[5] Both methods gave almost identical times of minimum. These are listed in Table 2, where the observed eclipses are labelled with the corresponding orbit number starting from 1.

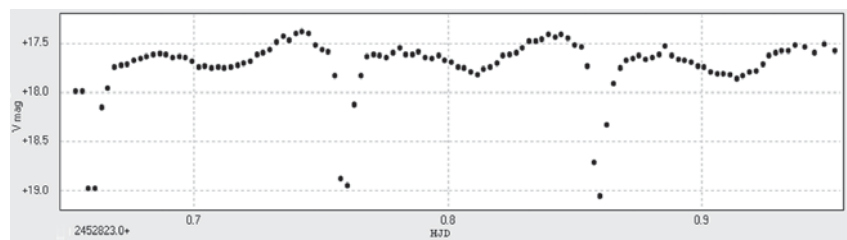

**Figure 1.** Lightcurve of SDSSJ1702 obtained on 2003 July 3 at NOFS *(Arne Henden)*.

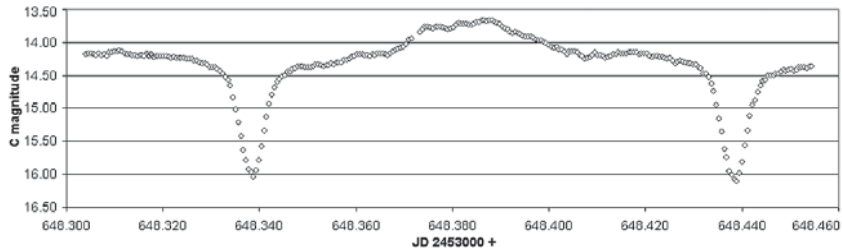

**Figure 2.** Lightcurve on 2005 October 4 *(David Boyd)*.

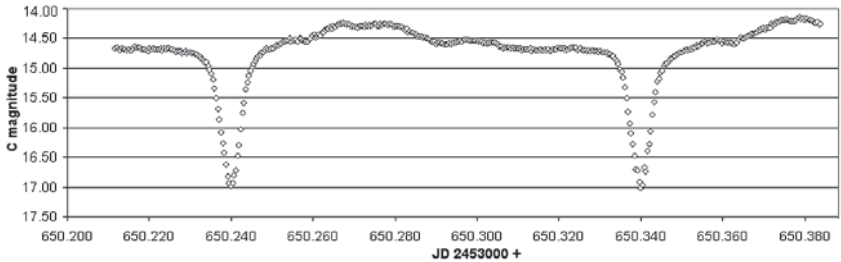

**Figure 3.** Lightcurve on 2005 October 6 *(Arto Oksanen)*.

The orbital period calculated by a linear fit to these times of minimum is

$P_{orb} = 0.1000800 \pm 0.0000012$d (2h 24m 6.92±0.11s)

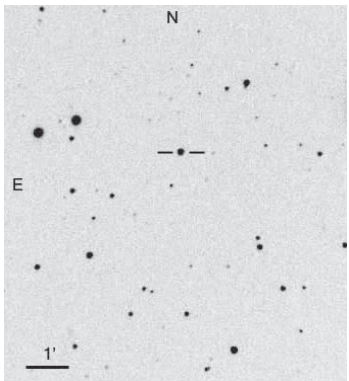

**Figure 4.** Field of SDSSJ1702.

This puts the system in the centre of the period gap. The residuals between the observed times of minimum and those calculated using this orbital period are shown in Figure 5.

Times of minimum were also measured for the three eclipses observed by Henden on 2003 July 3. These are listed in Table 3, together with the orbit numbers relative to those in Table 2 which best fit a linear extrapolation back in time using an orbital ephemeris calculated from the 2005 October data.

Predicted times for the 2003 July eclipses based on linear extrapolation using the 2005 October ephemeris are within 0.018d of the observed times, i.e. less than 20% of the orbital period. The uncertainty in this extrapolation at the one standard deviation level from the uncertainty in $P_{orb}$ is 0.01d. This level of agreement gives good confidence in the coherence of the 2003 and 2005 observations and the correctness of the orbit number assignment, assuming that the orbital period remained constant over the intervening two years.

Using the 3 eclipse times of minimum from 2003 and the 13 from 2005, recalculation of the orbital period yields a more precise value:

$P_{orb} = 0.10008215 \pm 0.00000001$d (2h 24m 7.098±0.001s)

The eclipse time of minimum ephemeris calculated using all 16 eclipse timings is

HJD = 2453648.236507(±0.000029) + 0.10008215(±0.00000001) * E

## Superhumps

All the 2005 October lightcurves show structure between the eclipses which could be due to superhumps. To determine whether these features are periodic, the eclipses were removed from the lightcurves and a period analysis carried out on the remaining 2005 October data using the ANOVA method in *Peranso*.[6] The resulting power spectrum is shown in Figure 6 and the region around the highest peak is expanded in Figure 7.

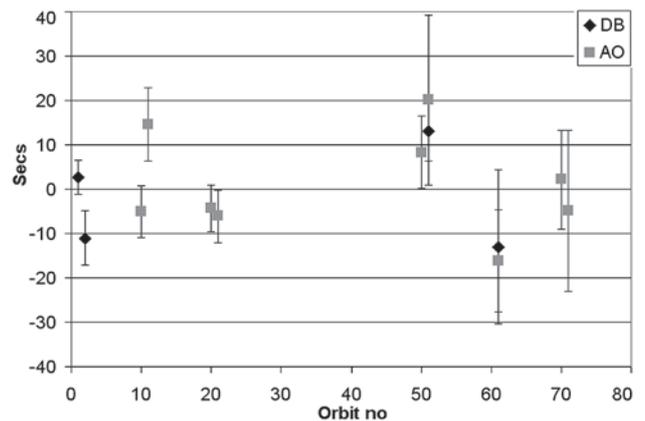

**Figure 5.** Residuals to the orbital period calculated using the data from 2005 October.

**Table 2. Eclipse times of minimum in 2005 October**

| Orbit no. | Time of minimum (HJD) | Error (d) |
|---|---|---|
| 1 | 2453648.33670 | 0.00005 |
| 2 | 2453648.43662 | 0.00007 |
| 10 | 2453649.23733 | 0.00007 |
| 11 | 2453649.33763 | 0.00010 |
| 20 | 2453650.23814 | 0.00006 |
| 21 | 2453650.33820 | 0.00007 |
| 50 | 2453653.24068 | 0.00010 |
| 51 | 2453653.34090 | 0.00022 |
| 51 | 2453653.34082 | 0.00008 |
| 61 | 2453654.34128 | 0.00013 |
| 61 | 2453654.34132 | 0.00020 |
| 70 | 2453655.24221 | 0.00013 |
| 71 | 2453655.34221 | 0.00021 |

**Table 3. Eclipse times of minimum in 2003 July**

| Orbit no. | Time of minimum (HJD) | Error (d) |
|---|---|---|
| −8239 | 2452823.65974 | 0.00011 |
| −8238 | 2452823.75953 | 0.00005 |
| −8237 | 2452823.85990 | 0.00004 |

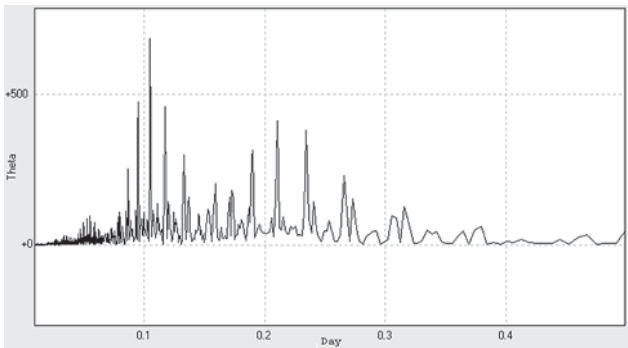

**Figure 6.** Power spectrum after removal of eclipses.

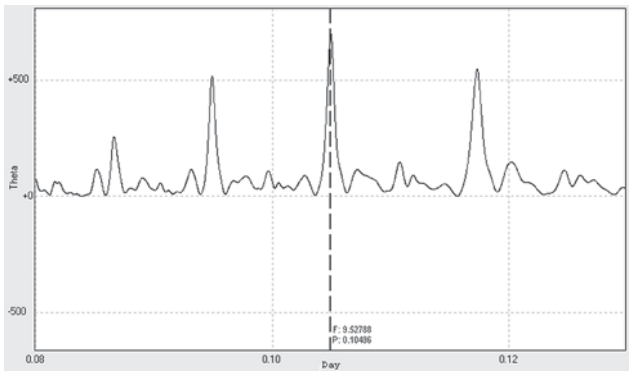

**Figure 7.** Power spectrum around the highest peak.

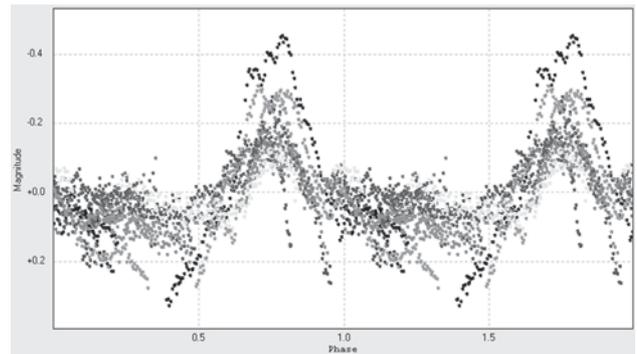

**Figure 8.** Phase diagram of the 2005 October data (excluding eclipses) folded on the superhump period.

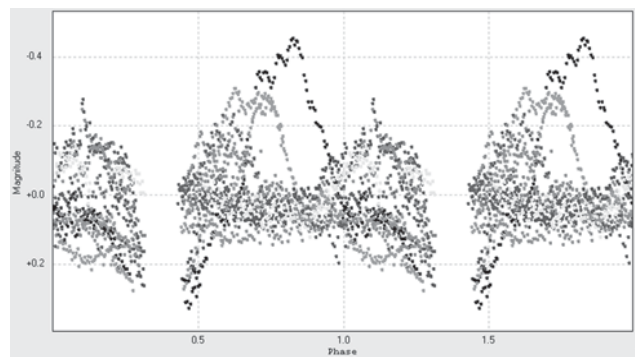

**Figure 9.** Phase diagram of the 2005 October data (excluding eclipses) folded on the orbital period.

The strongest signal in the data is at 0.10496±0.00015d and there are aliases at ±1c/d as expected. Lomb−Scargle and PDM analyses gave similar results. We interpret this signal as the superhump period. A phase diagram of the 2005 October data folded on this period is shown in Figure 8. To test the stability of this period during the outburst, the data were reanalysed in two groups, October 4−6 and October 9−11. Both groups gave results compatible with the above figure so we conclude that the superhump period did not change detectably during the outburst. A superhump period of 0.10496d represents a period excess of 4.9%. This value is consistent with the distribution of period excess vs orbital period given in Figure 6.6 in Hellier.[7]

As a further test of the reality of the superhump signal, the 2005 October data were folded on the orbital period. Figure 9 indicates that there is no consistent signal at the orbital period.

Humps are also present in the 2003 July observations of the system in quiescence shown in Figure 1. A period analysis of this data after removing the eclipses gives a hump period of 0.1004±0.0015d, indicating that these are normal orbital humps.

## Conclusion

These observations show that SDSS J170213.26+322954.1 is an eclipsing UGSU-type cataclysmic variable with an orbital period of 0.10008215±0.00000001d. During the superoutburst in 2005 October, it exhibited superhumps with a period of 0.10496±0.00015d representing a period excess of 4.9%.

## Acknowledgments

We thank the referee for helpful comments. AO thanks the American Association of Variable Star Observers (AAVSO) and the Curry Foundation for funding his CCD camera.

**Addresses: DB:** 5 Silver Lane, West Challow, Wantage, OX12 9TX, UK [drsboyd@dsl.pipex.com]
**AO:** Verkkoniementie 30, FI-40950 Muurame, Finland [arto.oksanen@jklsirius.fi]
**AH:** AAVSO, 25 Birch Street, Cambridge, MA 02138, USA [arne@aavso.org]